\documentclass[reprint,pra,showpacs,amsmath,amssymb,superscriptaddress,showkeys,longbibliography]{revtex4-1}

\usepackage{hyperref}
\usepackage{braket}
\usepackage{graphicx}
\usepackage{xcolor}

\begin{document}

\title{Wave-particle correlations in multiphoton resonances of coherent light-matter interaction}

\author{Th. K. Mavrogordatos}
\email[Email address: ]{themis.mavrogordatos@fysik.su.se}
\affiliation{Department of Physics, Stockholm University, SE-106 91, Stockholm, Sweden}
\affiliation{ICFO -- Institut de Ci\`{e}ncies Fot\`{o}niques, The Barcelona Institute of Science and Technology, 08860 Castelldefels (Barcelona), Spain}
 
\date{\today}
  
\begin{abstract}
We discuss the conditional measurement of field amplitudes by a nonclassical photon sequence in the Jaynes-Cummings (JC) model under multiphoton operation. We do so by employing a correlator of immediate experimental relevance to reveal a distinct nonclassical evolution in the spirit of [G. T. Foster {\it et al.}, Phys. Rev. Lett. {\bf 85} 3149 (2000)]. The correlator relies on the complementary nature of the pictures obtained from different unravelings of a JC source master equation. We demonstrate that direct photodetection entails a conditioned separation of timescales, a quantum beat and a semiclassical oscillation, produced by the coherent light-matter interaction in its strong-coupling limit. We single the quantum beat out in the analytical expression for the waiting-time distribution, pertaining to the particle nature of the scattered light, and find a negative spectrum of quadrature amplitude squeezing, characteristic of its wave nature for certain operation settings. Finally, we jointly detect the dual aspects through the wave-particle correlator, showing an asymmetric regression of fluctuations to the steady state which depends on the quadrature amplitude being measured.
\end{abstract}

\pacs{03.65.Yz, 32.80.-t, 42.50.Ar, 42.50.Lc, 42.50.Ct}
\keywords{Wave/particle duality, photon blockade, Jaynes-Cummings model, direct photodetection, waiting-time distribution, balanced homodyne detection, squeezed light, quantum interference}

\maketitle

\section{Introduction}

As early as in 1909, Planck suggested~\cite{Planck1909} that ``one should first try to move the whole difficulty of the quantum theory to the domain of the interaction of matter with radiation.'' The suggestion was followed up with enough rigour in the relevant proposal by Bohr, Kramers and Slater (BKS) put forward in 1924~\cite{BKS1924, Slater1924}. Although BKS did not mention light particles, the continuous wave amplitude of ambient radiation interacting with matter was to determine probabilities for discrete transitions between stationary states~\cite{Einstein1917} in the classical attitude of keeping waves and particles separate. Their approach foundered since it failed to causally connect the downward jump of an emitting atom to the subsequent upward jump of a particular absorbing atom, excluding direct correlation between individual quantum events and contradicting X-ray experiments~\cite{Bothe1925, Compton1925}. In other words, the proposal only allowed super-Poissonian photoelectron count fluctuations~\cite{CarmichaelTalk2000}. Any experimental quest for the disallowed correlations should then involve a method for engineering the fluctuations of light on the scale of Planck's energy quantum~\cite{CarmichaelQO2, CarmichaelTalk2000, Chang2014}.

Over the past two decades, the exquisite control acquired over cavity and circuit QED implementations has reappraised the vary nature of Bohr's~\cite{Bohr1913} indivisible quantum jump~\cite{Minev2019} alongside the manifestation of coherence in nonlinear optics at the level where adding (or subtracting) a single photon shifts the resonance frequencies with distinct observational consequences~\cite{Saito2004, Birnbaum2005, Faraon2008, Bishop2009, Fink2017, Xu2017, Hamsen2017, Pietikainen2019, Najer2019, Sett2022}. Nonlinearity is here to be understood in terms of multiphoton transitions~\cite{Zubairy1980, Munoz2014} -- the quantum mechanical substitute against the linear Schr\"{o}dinger equation -- taking place within an energy spectrum influenced by the light-matter coupling strength~\cite{CarmichaelQO1, CarmichaelQO2}. Two recent experiments studying multiphoton resonances~\cite{Hamsen2017, Najer2019} report on nonclassical features evinced by ensemble-averaged quantities, mainly revolving around second and higher-order correlation functions of the light scattered from an ion or a quantum dot. At the same time, histograms in the phase space depicting real time single-shot data of both quadratures of the transmitted output field~\cite{Sett2022} detail the breakdown of photon blockade. 

Conditioned balanced homodyne detection (BHD) has been proposed and implemented in~\cite{GiantViolations2000, Reiner2001} as an extension~\cite{Davies1969} of the intensity correlation technique introduced by Hanbury-Brown and Twiss in 1956~\cite{Brown1956,BrownHTwissI,BrownHTwissII,Mandel1958, Glauber1963I,Glauber1963II,OpenSystems2013}. The proposal unveils the tensions raised by wave/particle duality in a distinct manner~\cite{CarmichaelTalk2000}, by detecting light as particle and wave -- its both character roles: the measured wave property (radiation field amplitude) is correlated with the particle detection (photoelectric count). A certain amount of motivation in fusing the wave and particle ideas about light~\cite{Einstein1909} can be traced back to the prominent nonclassicality of resonance fluorescence~\cite{Carmichael1985, Rice88}, where we find squeezing~\cite{Walls1983} along the quadrature which is in phase with the mean scattered field amplitude~\cite{WallsZoller1981}. A negative value of the corresponding normal-ordered variance is precisely the source of antibunching~\cite{Kimble1977, Mandel1982} in the weak-field limit -- a direct association with the nonclassical statistics of a phased oscillator~\cite{BadCavityLimit, CarmichaelQO1}. More recently, the regression of the resonance fluorescence source field has been measured in the experiment of Ref.~\cite{Gerber2009}, while the amplitude-intensity correlations of selected transitions in a three-level ladder atom and a $V$-shaped atom are known to exhibit a characteristic temporal asymmetry~\cite{Xu2015}.

\begin{figure*}[t]
 \includegraphics[width=0.7\textwidth]{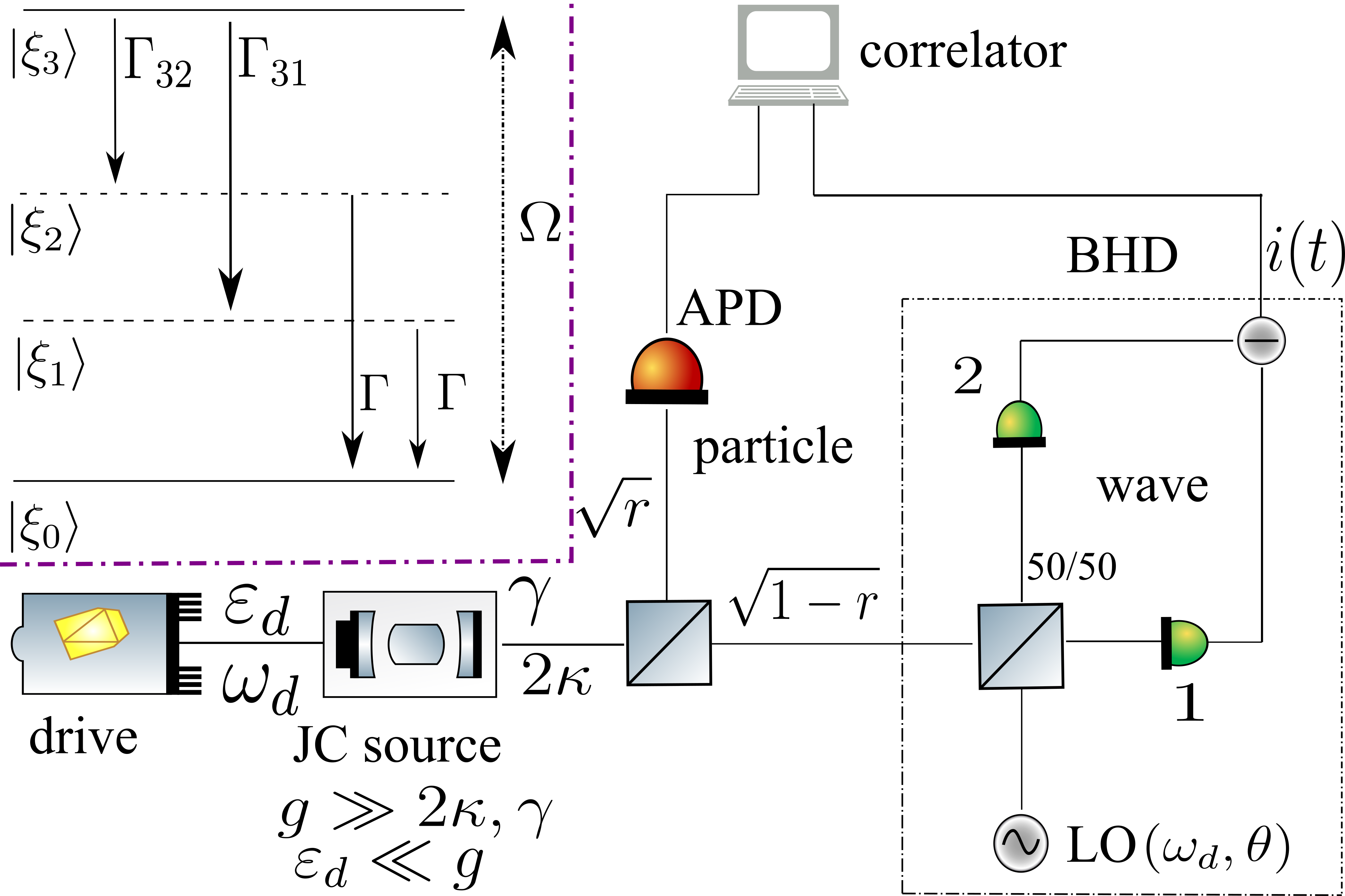}
 \caption{{\it Unraveling scheme and dynamical system response function.} Schematic diagram of the wave-particle correlator after~\cite{GiantViolations2000, CarmichaelFosterChapter}: A fraction $1-r$ of the input photon flux -- originating from the JC oscillator coherently driven at frequency $\omega_d$ and with strength $\varepsilon_d$ -- is directed towards a balanced homodyne detector (BHD) outputting a photocurrent $i(t)$, while the rest (fraction $r$) is sent to a photon counter -- an avalanche photodiode (APD) -- triggering the sampling of $i(t)$. The local oscillator (LO) phase is $\theta$ and its frequency is set to $\omega_d$. Changing the values of $r$ and $\theta$ leads to different ME unravelings. The upper left part depicts the four-state scheme used~\cite{Shamailov2010} to model a two-photon JC resonance in the secular approximation~\cite{Cohen_Tannoudji_1977}. The effective two-photon Rabi frequency is $\Omega=2\sqrt{2}\varepsilon_d^2/g$, while the transition rates $\Gamma_{3(2)1}$ and $\Gamma$ are linear combinations of the decoherence rates $2\kappa$ and $\gamma$~\cite{CarmichaelQO2, Shamailov2010}.}
 \label{fig:scheme}
\end{figure*}

How then does the quantum substitute to classical nonlinearity unfold in an explicitly open-system setup? To address the question on what information can we {\it operationally} read off from a quantum nonlinear source, in this report we bring up the ability of the wave-particle correlator to produce complementary unravelings of the source dynamics. When the source field is small and nonclassical, its fluctuations dominate over the steady-state amplitude -- see e.g., the giant violation of classical inequalities for squeezed light~\cite{GiantViolations2000, Foster2000}. In this work, we move to a largely unexplored region of cavity QED~\cite{Reiner2001, Denisov2002, CarmichaelFosterChapter}, one where intense quantum fluctuations produce a continual disagreement with the mean-field response of the source in what defines a so-called strong-coupling ``thermodynamic limit''~\cite{PhotonBlockade2015, SC2019}. Instead of focusing on a third-order correlation function by averaging over an ensemble of homodyne current samples to recover a signal out of shot noise, our interest is actually with what makes the homodyne current record in {\it individual realizations}. We refer to this decomposition as the {\it wave-particle correlator unraveling of the master equation (ME)} according to the scheme depicted in Fig~\ref{fig:scheme}, combining conditional measurements with quantum measurement~\cite{CarmichaelFosterChapter}. Light emanating from a coherently driven Jaynes-Cummings (JC) oscillator is split between two paths. Photodetections along one of the arms trigger homodyne measurements of the quadrature amplitudes, producing the current $i(t)$. 

Our narrative is structured as follows. In Sec.~\ref{sec:sourceME}, we formulate the driven dissipative {\it unconditional} dynamical description of an ensemble-averaged evolution. Section~\ref{sec:particleasp} is devoted to the derivation of an analytical expression for the waiting-time distribution of the forward-scattered photons in the limit of vanishing spontaneous emission for the two-state atom -- the so-called ``zero system-size'' limit of absorptive optical bistability -- which is of special interest to the persistence of photon blockade~\cite{PhotonBlockade2015, SC2019}. In Sec.~\ref{sec:waveasp}, we use the minimal four-state model of a cascaded two-photon resonance to derive analytical results for the transmission spectrum and for the spectrum of squeezing. Finally, in Sec.~\ref{sec:pwcorr}, we detail the unraveling scheme based on the wave-particle correlator applied to the light emanating from the JC oscillator under multiphoton resonance operation, before summarizing our main findings in the Conclusions. 

\section{The source master equation: dressing of dressed states in action}
\label{sec:sourceME}

Without reference to any particular unraveling strategy, the reduced density matrix of the open system evolves according to the standard Lindblad ME carrying the many pictures forward,
\begin{equation}\label{eq:ME1}
\begin{aligned}
 \frac{d\rho}{dt}\equiv \mathcal{L}\rho=&-i[\omega_0(\sigma_{+}\sigma_{-} + a^{\dagger}a)+g(a\sigma_{+}+a^{\dagger}\sigma_{-}),\rho]\\
 &-i[\varepsilon_d (a e^{i\omega_d t} + a^{\dagger}e^{-i\omega_d t}),\rho]\\
 &+\kappa (2 a \rho a^{\dagger} -a^{\dagger}a \rho - \rho a^{\dagger}a)\\
 &+\tfrac{\gamma}{2}(2\sigma_{-}\rho \sigma_{+} - \sigma_{+}\sigma_{-}\rho - \rho \sigma_{+}\sigma_{-}),
 \end{aligned}
\end{equation}
where the terms of the first two lines on the right-hand side correspond to the dynamical response function of the scattering center, a coherently driven cavity mode $a$ strongly and resonantly coupled to a two-state atom (of transition frequency $\omega_0$) with strength $g\ll \omega_0$~\cite{Burgarth2022oneboundtorulethem}, and the last two terms describe the two dissipative channels opened by the coupling of the oscillator to the environment -- photon loss from the cavity at rate $2\kappa$ and spontaneous emission from the two-state atom at rate $\gamma$. The experimenter controls the inputs to the JC oscillator, tuning the frequency $\omega_d$ of the drive and regulating its strength $\varepsilon_d$, while they read outputs treated as excitations of two independent zero-temperature reservoirs. Here we will be operating under the hierarchy of scales $g \gg (2\kappa, \gamma)$ and $\varepsilon_d/g \ll 1$, while there is a significant detuning between the drive field and cavity mode, $\Delta\omega_d\equiv(\omega_d-\omega_0) \sim g$ to allow for selectively exciting multiphoton resonances between the dressed states in the manifold: $|\xi_0\rangle=|0, -\rangle$ and $|\xi_{n/(n+1)}\rangle=\tfrac{1}{\sqrt{2}}(|n, -\rangle\mp|n-1,+\rangle)$ for $n\geq 1$, where $|n,\pm\rangle \equiv |n\rangle |\pm \rangle$, with $|\pm\rangle$ the upper and lower states of the two-state atom and $\ket{n}$ the Fock states of the cavity field. For the minimal model introduced in~\cite{Shamailov2010, Lledo2021}, the two-photon excitation occurs across the ground state $|\xi_0\rangle$ and $|\xi_3 \rangle$, while the cascaded decay is mediated by the first excited doublet states $|\xi_{1,2}\rangle$. The mediation is observed through a coherent quantum beat~\cite{Shamailov2010}, present in any single realization we will meet.
\begin{figure*}[t]
 \includegraphics[width=0.99\textwidth]{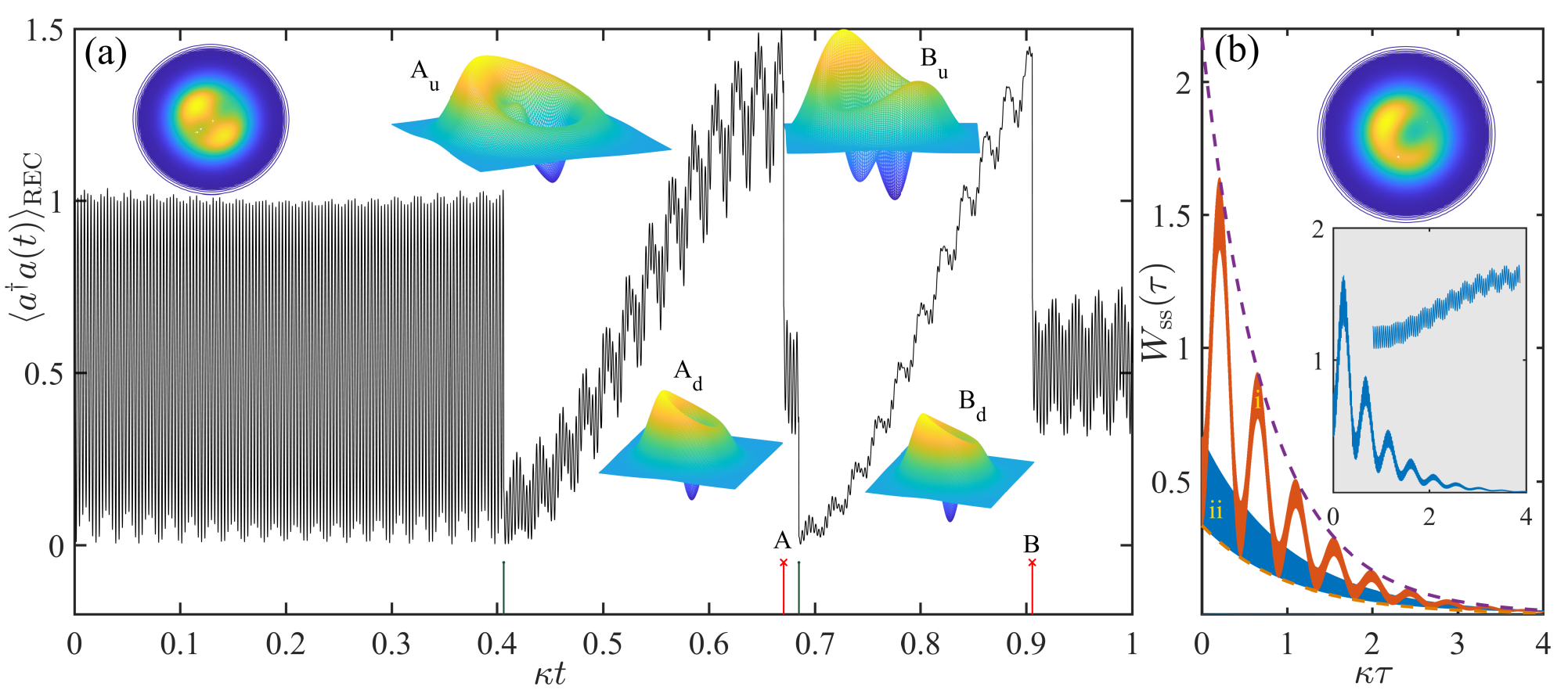}
 \caption{{\it Conditioned separation vs. unconditional coexistence of timescales in the particle aspect of light.} {\bf (a)} Sample trajectory of the conditioned intracavity photon number $\braket{a^{\dagger}a(t)}_{\rm REC}$ from the decomposition of the ME~\eqref{eq:ME1} under direct photodetection ($r=1$) with initial condition $\rho(0)=|1,-\rangle \langle 1, -|$. The green (red crossed) strokes underneath indicate spontaneous (cavity) emission events. The surface plots in frame (a) depict the transient conditioned Wigner function of the cavity field from the high to the lower photon excitation across the two cavity emissions marked A and B, while the contour plots in both frames depict the unconditioned Wigner function from the steady-state numerical solution of the ME~\eqref{eq:ME1} in \textsc{Matlab}'s Quantum Optics Toolbox~\cite{Tan1999}. {\bf (b)} Steady-state waiting-time distribution of the forward-scattered light, $W_{\rm ss}(\tau)\equiv \kappa^{-1} w_{{\rm ss},\rightarrow}(\tau)$ from Eq.~\eqref{eq:Wdist}, analytically derived from the minimal model and plotted for two different values of the driving strength such that: $\Omega/\kappa=10/\sqrt{2}$ for (i) [corresponding to the same drive strength as in (a)] and $1/\sqrt{2}$ for (ii). The lower inset depicts the numerically determined $W_{\rm ss}(\tau)$ from the quantum regression formula, alongside the superposition of the quantum beat to the first half period of the Rabi oscillation. In frame [(a), (b)], $\gamma/(2\kappa)=[1,0]$, while in both frames $g/\kappa=1,000$. For (a), a Monte-Carlo (MC) algorithm was developed {\it ad hoc} in \textsc{Matlab} using a basis truncated at the 14-photon level.}
 \label{fig:particleWss}
\end{figure*}
To derive analytical results for two-time averages in the steady state we employ the minimal four-state model and the quantum-regression formula~\cite{CarmichaelQO1} under the secular approximation~\cite{Cohen_Tannoudji_1977} and in the strong-coupling limit of nonperturbative QED ($g \gg \kappa, \gamma/2$). Adiabatically eliminating the intermediate states $|\xi_1 \rangle$ and $|\xi_2\rangle$ leads to the following effective ME [Eq. (18) of Ref.~\cite{Shamailov2010}]
\begin{equation}\label{eq:ME2}
\begin{aligned}
  &\frac{d\rho}{dt}=\tilde{\mathcal{L}}\rho \equiv -(i/\hbar)[\tilde{H}_{\rm eff},\rho]+\Gamma_{32} \mathcal{D}[|\xi_2\rangle \langle \xi_3|](\rho)\\
  &+ \Gamma_{31} \mathcal{D}[|\xi_1\rangle \langle \xi_3|](\rho)
  + \Gamma \mathcal{D}[|\xi_0\rangle \langle \xi_1|](\rho) + \Gamma \mathcal{D}[|\xi_0\rangle \langle \xi_2|](\rho),  
  \end{aligned}
\end{equation}
with an effective Hamiltonian modeling the driving of a two-photon transition,
\begin{equation}
 \tilde{H}_{\rm eff}\equiv\sum_{k=0}^{3} \tilde{E}_{k} |\xi_k\rangle \langle \xi_k| + \hbar \Omega (e^{2i\omega_d t} |\xi_0\rangle \langle \xi_3| + e^{-2i\omega_d t} |\xi_3\rangle \langle \xi_0|).
\end{equation}
The intermediate states are not to be discarded since they take part in the cascaded process. The shifted energy levels dressed by the drive as second-order corrections in $\varepsilon_d/g$ are
\begin{subequations}\label{eq:shifts}
\begin{align}
 &\tilde{E}_0=E_0 + \hbar\delta_0(\varepsilon_d) = \hbar \sqrt{2} \varepsilon_d^2/g, \\
 & \tilde{E}_1=E_1 + \hbar\delta_1(\varepsilon_d) = \hbar \{\omega_0 -  g - [(20 + 19\sqrt{2})/7]\varepsilon_d^2/g\}, \\
 & \tilde{E}_2=E_2 + \hbar\delta_2(\varepsilon_d) = \hbar \{\omega_0 +  g +  [(20 - 19\sqrt{2})/7]\varepsilon_d^2/g\}, \\
 & \tilde{E}_3=E_3 + \hbar\delta_3(\varepsilon_d) =  \hbar (2\omega_0 - \sqrt{2}  g -  \sqrt{2}\, \varepsilon_d^2/g),
\end{align}
\end{subequations}
while the effective two-photon Rabi frequency is $\Omega=2\sqrt{2}\, \varepsilon_d^2/g$~\cite{Lledo2021}. In the effective ME~\eqref{eq:ME2}, we define $\mathcal{D}[X](\rho)\equiv X\rho X^{\dagger}-(1/2)\{X^{\dagger}X, \rho\}$, while the transition rates between the four levels, in the special case where $\gamma=2\kappa$, are~\cite{Shamailov2010}
\begin{subequations}
\begin{align}
 &\Gamma_{31}\equiv\frac{\gamma}{4}+(\sqrt{2}+1)^2 \frac{\kappa}{2}=\frac{\gamma}{4}[1+(\sqrt{2}+1)^2], \label{eq:Gamma31} \\
 & \Gamma_{32}\equiv\frac{\gamma}{4}+(\sqrt{2}-1)^2 \frac{\kappa}{2}=\frac{\gamma}{4}[1+(\sqrt{2}-1)^2], \\
 & \Gamma\equiv\frac{\gamma}{2}+\kappa=\gamma.
\end{align}
\end{subequations}
Including perturbative corrections, the two-photon resonance must be excited with a drive frequency $\omega_d$ obeying $2\omega_d=(\tilde{E}_3-\tilde{E}_1)/\hbar=2\omega_0-\sqrt{2}g + \delta_3(\varepsilon_d)-\delta_0(\varepsilon_d)$, yielding $\Delta\omega_d\equiv \omega_d-\omega_0=-g/\sqrt{2} - \sqrt{2}\, \varepsilon_d^2/g$. The effective ME~\eqref{eq:ME2} governs the evolution of the matrix elements in the dressed-state basis and needs to be solved twice~\cite{CarmichaelQO1}; first to obtain the steady-state density matrix $\rho_{\rm ss}$ and second to advance the argument of the evolution up to the time $\tau$. 
\begin{figure*}[t]
\centering
 \includegraphics[width=0.95\textwidth]{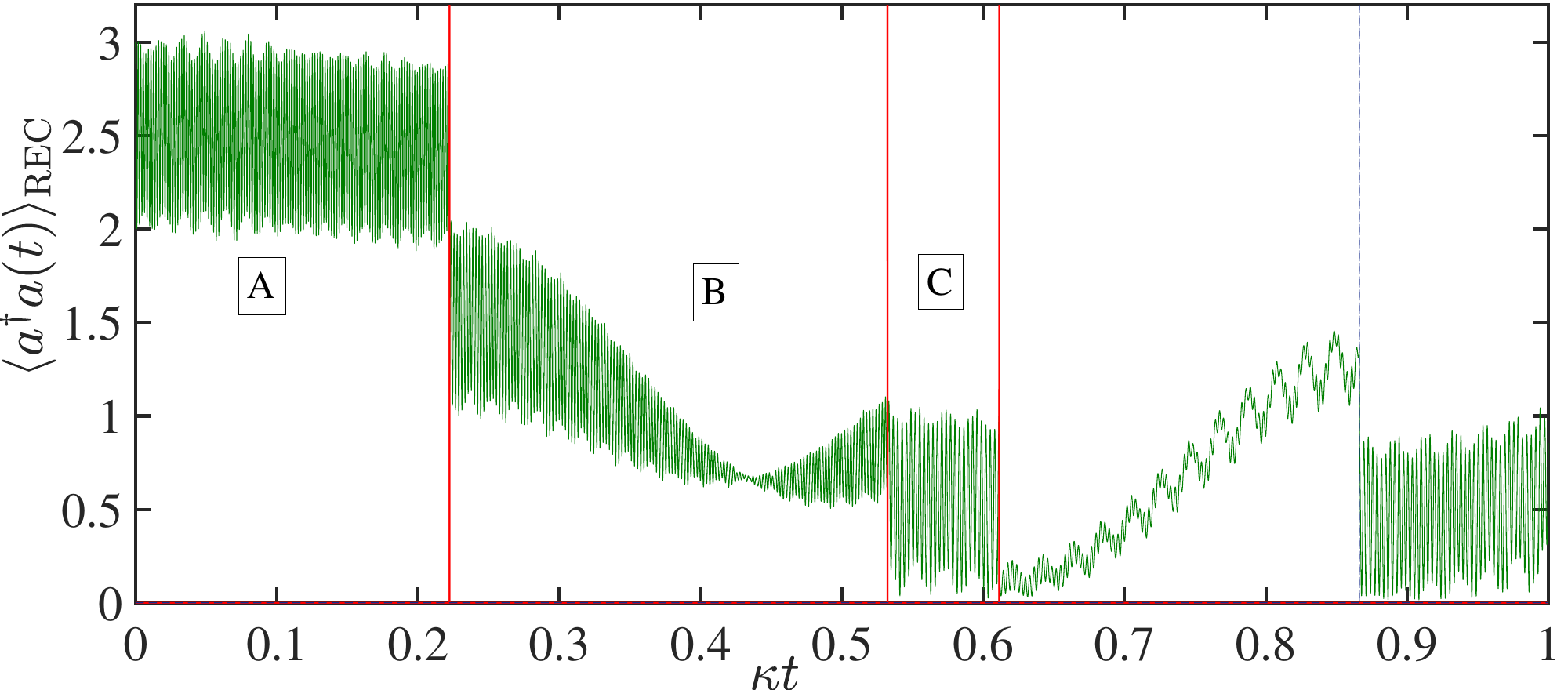}
 \caption{{\it Conditional quantum beat frequencies in the transient.} Sample trajectory of the conditional cavity photon number under direct photodetection for the same parameters used in Fig.~\ref{fig:particleWss}(a) except for the initial pure state which is now set to $\rho(0)=|3,- \rangle \langle 3,-|$. The dominant quantum beat (qb) frequency of oscillation in the three segments A, B, C of the evolution is $\omega_{qb}=2\sqrt{3}g, 2\sqrt{2}g, 2g$, respectively. Red solid lines mark the cavity emissions and the dashed blue line indicates a spontaneous emission. A Hilbert space truncated at the 25-photon level was used in the Monte-Carlo algorithm.} 
 \label{fig:qb}
\end{figure*}

Three timescales revealed by the minimal model are present in the unraveling of Fig.~\ref{fig:particleWss}(a) under direct photodetection as well as in the numerically obtained waiting-time distribution of Fig.~\ref{fig:particleWss}(a). The fastest of these three corresponds to the quantum beat, with frequency equal to $\nu\equiv (\tilde{E}_2-\tilde{E}_1)/\hbar=2g + \delta_2-\delta_1=2g + \mathcal{O}(\varepsilon_d^2/g)$. An intermediate timescale, which is missed in the secular approximation and therefore in the analytical treatment, is present in the numerically-extracted evolution due to the asymmetry between the excitation paths $|\xi_3\rangle \to |\xi_1\rangle \to |\xi_0\rangle$ and $|\xi_3\rangle \to |\xi_2\rangle \to |\xi_0\rangle$, since $\Gamma_{31}/\Gamma_{32}\approx 5.8$ for $\gamma/(2\kappa)=0$ (see Sec. 3.2 of~\cite{Shamailov2010} and, {\it e.g.}, the field-amplitude oscillations in the top panel of Fig.~\ref{fig:orthogonal}). In a frame rotating with the drive and for $\Delta\omega_d=-g/\sqrt{2}$, the unperturbed energies of the intermediate states are $E^{(0)}_{1,2}=g(1\mp \sqrt{2})/\sqrt{2}$ whereas the outer states $|\xi_3 \rangle$ and $|\xi_0\rangle$ have zero energy. It follows that the excitation path $|\xi_3\rangle \to |\xi_1\rangle \to |\xi_0\rangle$ is associated with a periodic structure including $2\sqrt{2}/(\sqrt{2}-1)\approx 6.8$ quantum beat cycles while the path $|\xi_3\rangle \to |\xi_2\rangle \to |\xi_0\rangle$ is associated with a period of about $1.2$ times the beat cycle, manifested as an occasional deviation from the pattern dictated by the dominant path. The slowest timescale corresponds to the semiclassical oscillation with frequency $2\Omega=4\sqrt{2}\varepsilon_d^2/g$~\cite{Lledo2021} and is associated with driving a saturable (effective) two-level transition, similar to ordinary resonance fluorescence.

Numerical results show that initializing the JC oscillator at a state $|n,-\rangle$ with high excitation $n$ outside the manifold of the minimal model devised to describe the steady state, creates a sequence of {\it transient} quantum beats in the conditional emission rate $2\kappa\langle a^{\dagger}a(t) \rangle_{\rm REC}$, with decreasing frequencies $2\sqrt{n}g, 2\sqrt{n-1}g,\ldots$ as we move down the JC ladder in response to successive quantum jumps. The initial condition chosen in Fig.~\ref{fig:qb} activates a quantum beat formed between the dressed states $|\xi_5\rangle$ and $|\xi_6\rangle$ which is interrupted by the first jump at $\kappa t_1 \approx 0.22$. Another quantum beat of decreased frequency develops involving the couplet states $|\xi_4\rangle$ and $|\xi_3\rangle$, intertwined with an emerging semiclassical oscillation; this pattern is again interrupted at $\kappa t_2 \approx 0.53$. Segment C corresponds to a quantum beat of frequency $2g$, the one accounted for by the minimal model and involving the first-excited couplet states $|\xi_2\rangle$ and $|\xi_1\rangle$. The evolution from that point onwards resembles the conditional separation of timescales shown in Fig.~\ref{fig:particleWss}(a).

\section{Particle aspect: intensity correlation and waiting-time distribution}
\label{sec:particleasp}

Armed with these observations regarding the timescales involved in the coherent part of the dynamical evolution, we activate only one arm of the wave-particle correlator to formulate our first unraveling of the source master equation. Furthermore, from now on we operate in the so-called limit of ``zero system size'' $\gamma/(2\kappa)\to 0$~\cite{SavageCarmichael1988, CarmichaelQO2} in which the length of the Bloch vector is preserved during the evolution, while the system-size scale parameter $n_{\rm sc, w}=[\gamma/(2\sqrt{2}g)]^2$ akin to a weak-coupling ``thermodynamic limit'' gives its place to $n_{\rm sc, s}=[g/(2\kappa)]^2$, whose divergence defines a ``strong-coupling'' limit~\cite{PhotonBlockade2015}.

We start by setting $r=1$. Figure~\ref{fig:particleWss}, from which we deduced the relevant timescales, is now seen from a different perspective: it introduces us to the main departure from the {\it unconditional} ME dynamics, namely to the theme of a conditioned separation between a fast and a slow timescale, precipitated in this particular case by spontaneous emissions. Preparing the JC oscillator in a pure state, in particular in $|n=1, -\rangle$ -- an equal-weight superposition of the dressed states $|\xi_1\rangle$ and $|\xi_2\rangle$ -- generates a coherent evolution dominated by a quantum beat of frequency $\nu \approx 2g$ until the first quantum jump. Such a collapse marks the onset of significantly slower semiclassical Rabi oscillations of frequency $2\Omega=4\sqrt{2}\varepsilon_d^2/g \ll 2g$, a feature familiar to us from the saturation of a two-level transition~\cite{CarmichaelQO1, Tian1992}. An intermediate timescale, of a period during which about seven quantum beat cycles are completed, is also present owing to the asymmetry between the excitation paths $|\xi_3\rangle \to |\xi_1\rangle \to |\xi_0\rangle$ and $|\xi_3\rangle \to |\xi_2\rangle \to |\xi_0\rangle$~\cite{Shamailov2010}. This trend is interrupted by a cavity emission of the second photon from the pair, bringing the cavity mode to a lower-excitation state and reviving the quantum beat for a short while. Conditioned phase-space distributions of the cavity field at the two ends of the jump show a transition from an odd-parity superposition to the Wigner function of a single-photon-added coherent state (SPACS), evidence of the ``smooth transition between the particle-like and wavelike behavior of light''~\cite{Zavatta2004}. Note the reliance on a conditional measurement, the central element underpinning the generation of the quantum trajectory depicted in Fig~\ref{fig:particleWss}: in the experiment of~\cite{Zavatta2004} a seed coherent field is injected into the signal mode of a parametric amplifier, and the conditional preparation of the SPACS takes place every time a single photon is detected in the correlated idler mode.

Let us keep following further the trajectory generated by a direct-photodetection unraveling of the ME. The emission of the first photon from the second pair reinitiates the semiclassical oscillation, but now one where the superimposed quantum beat does not feature as prominently. This is due to the weaker participation of the states $|\xi_1 \rangle$ and $|\xi_2 \rangle$ in the conditioned state resulting after the second spontaneous emission event. The second recorded photon emission in the forward direction follows an even-state superposition in the cavity-field distribution, which is again reduced to a conditioned state with the Wigner function of a SPACS. Both quantum-superposition states in the distributions at the two ends of the jump are framed by remnants of steady-state bimodality.   

\begin{figure*}[t]
 \includegraphics[width=1.02\textwidth]{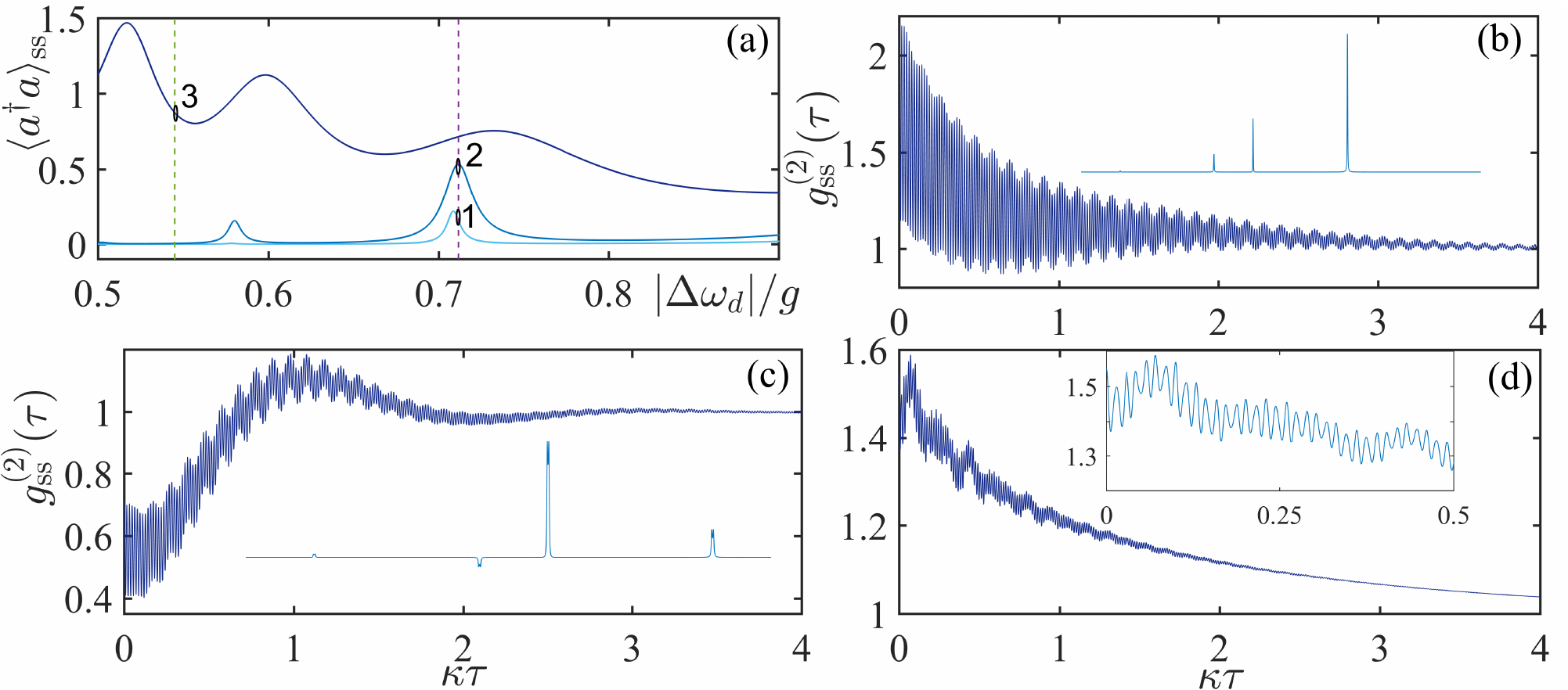}
 \caption{{\it Photon blockade and intensity correlation functions.} {\bf (a)} Steady-state intracavity photon number obtained from the numerical solution to the ME~\eqref{eq:ME1}, plotted against the (absolute value of the) scaled drive detuning $|\Delta \omega_d|/g$ for three increasing values of the drive strength $\varepsilon_d/g$: $0.03, 0.055$ and $0.16$ in one-to-one correspondence to the three resonance sequences of growing excitation in the cavity. Intersection 1 corresponds to the operation point selected for Fig.~\ref{fig:sampling}(a), 2 to Figs.~\ref{fig:sampling}(b, c), and 3 to Fig.~\ref{fig:sampling}(d). The three interaction points correspond to the inputs for the wave-particle correlator whose outputs are depicted in Fig.~\ref{fig:sampling}. The steady-state intensity correlation function of the forward-scattered photons $g_{\rm ss}^{(2)}(\tau)$ corresponding to these three operation points are plotted in frames {\bf (b), (c), (d)} respectively. The two insets in frames {\bf (b), (c)} depict the unlabeled spectrum of squeezing $\mathcal{S}_{\rightarrow}^{\theta}(\omega)$ against $(\omega-\omega_0)/\kappa$, calculated from Eq.~\eqref{eq:spsq} and the same conditions as in Figs.~\ref{fig:sampling}(a) [for $\theta=3\pi/4$] and (c) [for $\theta=\pi/4$], respectively . The inset in frame (d) focuses on the initial repetitive pattern formed by the superposition of quantum beats corresponding to Fig.~\ref{fig:sampling}(d). The intensity correlation functions have been numerically calculated in \textsc{Matlab}'s Quantum Optics Toolbox~\cite{Tan1999} using the quantum regression formula~\cite{CarmichaelQO1} in a Hilbert space of 14 photon levels.}
 \label{fig:g2s}
\end{figure*}

To assess the statistics of the light field stemming from its corpuscular nature, we have derived the waiting-time distribution of the forward-scattered radiation~\cite{Carmichael1989, Carmichael1993, Brandes2008} as (see Appendix A),
\begin{equation}\label{eq:Wdist}
 \begin{aligned}
  w_{{\rm ss},\rightarrow}(\tau)/(2\kappa)=&\tfrac{1}{2}\left[\tfrac{1}{2}e^{-\kappa \tau} + \tfrac{1}{6}e^{-\kappa \tau}\cos(\nu \tau)\right]\\
  &+\tfrac{3\Omega^2 e^{-3\kappa \tau/2}}{9\kappa^2-16\Omega^2}\sinh^2\left(\tfrac{\tau \sqrt{9\kappa^2-9\Omega^2}}{4}\right),
  \end{aligned}
\end{equation}
an expression plotted in Fig.~\ref{fig:particleWss}(b). The time $\tau$ waited between successive photoelectric emissions detected with perfect efficiency follows the superposition of the quantum beat [term $\propto \cos(\nu \tau)$] onto Rabi oscillations, the latter being developed as a characteristic ringing for increasing drive strength [last term in Eq.~\eqref{eq:Wdist}]. In contrast to ordinary resonance fluorescence, the distribution minima do not reach zero but are rather bound by the asymptote $\kappa \frac{1}{3} e^{-\kappa \tau}$ defined by the occupation of the intermediate levels in the cascaded process. The higher bound constraining the superposition of oscillations is reached in the limit $\varepsilon_d/\kappa \gg 1$. Figure~\ref{fig:particleWss}(b) shows very good agreement between Eq.~\eqref{eq:Wdist} and the numerical solution of the ME~\eqref{eq:ME1} combined with applying the quantum regression formula.   

\section{Wave aspect: the spectrum of squeezing}
\label{sec:waveasp}

Setting $r=0$ cancels the triggering in the wave-particle correlator and actuates solely the BHD arm, typically used for the measurement of amplitude-squeezing based on the ability of light to interfere with a strong coherent local oscillator (LO) field. These signatures speak of the continuous (wave) character of light. The spectrum of squeezing can be calculated by constructing the photocurrent autocorrelation $\overline{i(t)i(t+\tau)}$ for the selected quadrature of the field $A_\theta\equiv \frac{1}{2}(a e^{-i\theta}+a^{\dagger}e^{i\theta})$ as a time average over ergodic records~\cite{Cresser2001}, and taking the Fourier transform~\cite{CarmichaelQO2}.

Using once again the quantum regression formula we have derived analytical expressions for the normal (denoted by $:\,:$) and time-ordered variance $\langle:A_{\theta}(0)A_{\theta}(\tau):\rangle_{\rm ss}$ for two-photon resonance conditions (see Appendix B). For very low intracavity photon numbers, two of the peaks in the Fourier cosine transform (those corresponding to transitions between the ground state and the first-excited couplet) turn negative at low values of drive $\varepsilon_d/\kappa$, an effect which is maximized in the direction $\theta=\pi/4$. This direction is orthogonal to the anti-squeezed quadrature along which steady-state bimodality develops for stronger drive fields [see Fig.~\ref{fig:sampling}(a) and~\cite{Wigner2PB}]. Examples of a negative spectrum of squeezing are given in Fig.~\ref{fig:g2s} and in the top frame of Fig.~\ref{fig:orthogonal} in different regimes of the JC nonlinearity.  

Let us now revert to unconditional dynamics obeying the master equation~\eqref{eq:ME1} with reference to the nonclassical attributes of the particle and wave aspects {\it separately}, each revealed by a suitable multi-time correlation function. Figure~\ref{fig:g2s} presents the complementary nonexclusive probability densities expressed through the normalized second-order correlation function~\cite{Carmichael1993} for the forward-scattered field. We witness the transition from photon bunching to antibunching as steady-state bimodality builds up alongside a more pronounced Rabi splitting. Consequences of this change in photon statistics are revealed in the individual quantum trajectories of Figs.~\ref{fig:sampling}(a ,b), where the two occurring photon emissions are being spaced further apart in the course of an unraveling which combines both particle and wave aspects of light. When we move away from the two-photon resonance peak we come across a superposition of quantum beats with different frequencies and smaller amplitudes [see Fig~\ref{fig:g2s}(d)], a consequence of significant detuning of the drive field from an effective two-level resonance structure formed between the dressed JC eigenstates. 

\section{Particles triggering the sampling of waves}
\label{sec:pwcorr}

At this stage, let us put both arms in operation by setting $r=1/2$ to produce a sampling of a nonlinear stochastic Schr\"{o}dinger evolution solved by the conditioned wavefunction $|\psi_c(t)\rangle$. Our interest is with the full function of the wave-particle correlator, with reference to Fig.~\ref{fig:scheme}, in its ability to jointly detect the dual aspects of the incoming radiation. The set $(r, \theta)$ determines each one of the infinite unravelings of the ME: $0<r<1$, while the phase $\theta$ of the LO selects the quadrature of interest. The negative spectrum of squeezing we previously depicted for the two-photon resonance driving of the JC oscillator signifies a redistribution of fluctuations between the different quadratures of the cavity field. Here we aim to analyze the quantum fluctuations of the field as ostensible deviations from the steady state, combining conditional measurements with the process of quantum measurement.

We write the master equation for the source -- a driven JC oscillator in its zero ``system size limit'' $\gamma/(2\kappa)\to 0$ -- extended now to include the LO mode tuned to the frequency of the coherent drive $\omega_d$. The density matrix $\tilde{\rho}$ of the composite system in a frame rotating with $\omega_d$ reads~\cite{Reiner2001}:
\begin{equation}\label{eq:MEcorr}
\begin{aligned}
 \frac{d{\tilde{\rho}}}{dt}=&-i[-\Delta\omega_d(\sigma_{+}\sigma_{-} + a^{\dagger}a)+g(a\sigma_{+}+a^{\dagger}\sigma_{-}),\tilde{\rho}]\\
 &-i[\varepsilon_d (a + a^{\dagger}),\tilde{\rho}]+\kappa (2 a \tilde{\rho} a^{\dagger} -a^{\dagger}a \tilde{\rho} - \tilde{\rho} a^{\dagger}a)\\
 &+\kappa_{\rm LO}(2c \tilde{\rho} c^{\dagger}-c^{\dagger}c\tilde{\rho}-\tilde{\rho}c^{\dagger}c) + \kappa_{\rm LO} \beta[c^{\dagger}-c,\tilde{\rho}],
 \end{aligned}
\end{equation}
where $c$ and $c^{\dagger}$ are the raising and lowering operators for the LO driven mode with excitation amplitude $\beta$. Under the BHD scheme of Fig.~\ref{fig:scheme}, the total measured fields at the photodetectors 1 and 2 (in photon flux units) read
\begin{equation}\label{eq:fielddet}
 \mathcal{E}_{{\rm BHD}_{1,2}}=\pm i \sqrt{\kappa_{\rm LO}}\,c + \sqrt{\kappa(1-r)}\,a,
\end{equation}
and the field measured at the photon counting detector is
\begin{equation}
 \mathcal{E}_{\rm count}=\sqrt{2\kappa r}\, a.
\end{equation}
Under the assumption that the LO mode is unaffected by the signal mode, the density matrix factorizes into a piece corresponding to the driven JC oscillator under photon blockade, and a piece corresponding to the LO, $\tilde{\rho}=\tilde{\rho}_s |\beta\rangle \langle \beta|$.
Defining the LO flux as $f\equiv \kappa_{\rm LO}|\beta|^2$, we construct the following collapse super-operators,
\begin{widetext}
\begin{subequations}\label{eq:supops}
 \begin{align}
\mathcal{S}_{{\rm BHD}_{1,2}}\tilde{\rho}_s&=(\pm \sqrt{f} e^{i\theta} + \sqrt{2\kappa(1-r)}\,a)\,\tilde{\rho}_s\,(\pm \sqrt{f} e^{-i\theta} + \sqrt{2\kappa(1-r)}\,a^{\dagger})\label{eq:supopsBHD},\\
\mathcal{S}_{\rm count}\tilde{\rho}_s&=2\kappa r\, a \tilde{\rho}_s a^{\dagger} \label{eq:supopscounts}.
 \end{align}
\end{subequations}
\end{widetext}
Subtracting the action of super-operators~\eqref{eq:supops} from the Liouvillian of Eq.~\eqref{eq:MEcorr}, we arrive at an expression for the super-operator which governs the evolution of the un-normalized conditioned density operator $\tilde{\rho}_s (t)$,
\begin{widetext}
\begin{equation}\label{eq:sup2det}
(\mathcal{L}-\mathcal{S}_{{\rm BHD}_{1,2}}-\mathcal{S}_{\rm count})\tilde{\rho}_s=-i[-\Delta\omega_d(\sigma_{+}\sigma_{-} + a^{\dagger}a)+g(a\sigma_{+}+a^{\dagger}\sigma_{-})+\varepsilon_d(a+a^{\dagger}),\tilde{\rho}_s]-\kappa(a^{\dagger}a \tilde{\rho}_s + \tilde{\rho}_s a^{\dagger}a) - f\tilde{\rho}_s.
\end{equation}
\end{widetext}
Equation~\eqref{eq:sup2det} defines a non-Hermitian Hamiltonian which propagates the conditioned wavefunction $|\psi_c(t)\rangle$ of the system between measurements whose action is defined by the super-operators~\eqref{eq:supops}.

There are two types of detections which take place under the {\it conditioned} balanced homodyne detection (BHD) measurement scheme. The first comes from Eq.~\eqref{eq:supopscounts} and corresponds to detection ``clicks'' at the avalanche photodiode (APD). These collapses occur with probability $2\kappa r \langle \psi_c(t)|a^{\dagger}a|\psi_c(t)\rangle dt$. The second type of collapses concerns the ``clicks'' registered through the BHD. In any realistic homodyne measurement the LO photon flux $f$ is many orders of magnitude larger than the signal flux, here $2\kappa (1-r) \langle \psi_c(t)|a^{\dagger}a|\psi_c(t)\rangle$. This means that under the action of~\eqref{eq:fielddet}, a photoelectric emission corresponds with high probability to an annihilation of a LO photon. There is only a small probability $~f/[2\kappa \langle \psi_c(t)|a^{\dagger}a|\psi_c(t)\rangle]$ that a photon originated from the JC oscillator. We should note here that these two possibilities exist as a superposition and not as a classical choice, {\it either} one {\it or} the other. Despite the fact that the collapses are very small, on the characteristic timescale $(2\kappa)^{-1}$ for fluctuations in the signal field they occur very often. The quantum mapping into a stochastic differential equation -- a Schr\"{o}dinger equation with a stochastic non-Hermitian Hamiltonian -- involves a coarse graining in time, an expansion of the non-unitary evolution in powers of $\sqrt{2\kappa(1-r)/f}$ and the consideration of a stochastic process for the number of the emitted photoelectrons which depends on a conditioned wavefunction that satisfies the very same Schr\"{o}dinger equation. Taking that path leads to the following expressions for the difference in photocurrents between detectors 1 and 2 (in Fig.~\ref{fig:scheme}), and the wavefunction propagation between photodetections at the APD:
\begin{widetext}
\begin{equation}\label{eq:di}
di=-B (i\,dt - \sqrt{8\kappa(1-r)}\langle A_{\theta} \rangle_c\,dt - dW_t),
\end{equation}
\begin{equation}\label{eq:dpsi}
 d|\overline{\psi}_c\rangle=\bigg[\frac{1}{i\hbar}H^{\prime}\,dt + \sqrt{2\kappa(1-r)}\, a e^{-i\theta}(\sqrt{8\kappa(1-r)}\langle A_{\theta}(t) \rangle_c\,dt + dW_t)\bigg]|\overline{\psi}_c\rangle.
\end{equation}
\end{widetext}
Here $|\overline{\psi}_c\rangle$ is the un-normalized wavefunction and $H^{\prime}$ is the non-Hermitian Hamiltonian
\begin{equation}
\begin{aligned}
 H^{\prime}\equiv& H_{JC}-i\hbar \kappa a^{\dagger}a\\
 =&-\hbar \Delta\omega_d (a^{\dagger}a + \sigma_+ \sigma_-)+ \hbar g (a^{\dagger}\sigma_- + a \sigma_+)\\
 &+\hbar\varepsilon_d(a+a^{\dagger})-i\hbar \kappa a^{\dagger}a.
 \end{aligned}
 \end{equation}
In Eqs.~\eqref{eq:di} and~\eqref{eq:dpsi}, $\langle A_{\theta}(t) \rangle_c$ is the conditioned average of the quadrature amplitude selected by LO, $ \langle A_{\theta}(t) \rangle_c \equiv \tfrac{1}{2} \langle \psi_c(t)|e^{i\theta}a^{\dagger} + a e^{-i\theta}|\psi_c(t) \rangle$,
calculated with the normalized conditioned wavefunction $|\psi_c (t)\rangle$, $B$ is the detection bandwidth and $dW_t$ is the Wiener noise increment, the same in both equations~\eqref{eq:di},~\eqref{eq:dpsi}. In the Monte-Carlo algorithm developed to simulate the wave-particle correlator unravelling, the stochastic differential equation~\eqref{eq:dpsi} is solved by means of an explicit order 2.0 weak scheme proposed by Kloeden and Platen~\cite{KloedenPlaten}.
\begin{figure*}[ht]
 \includegraphics[width=\textwidth]{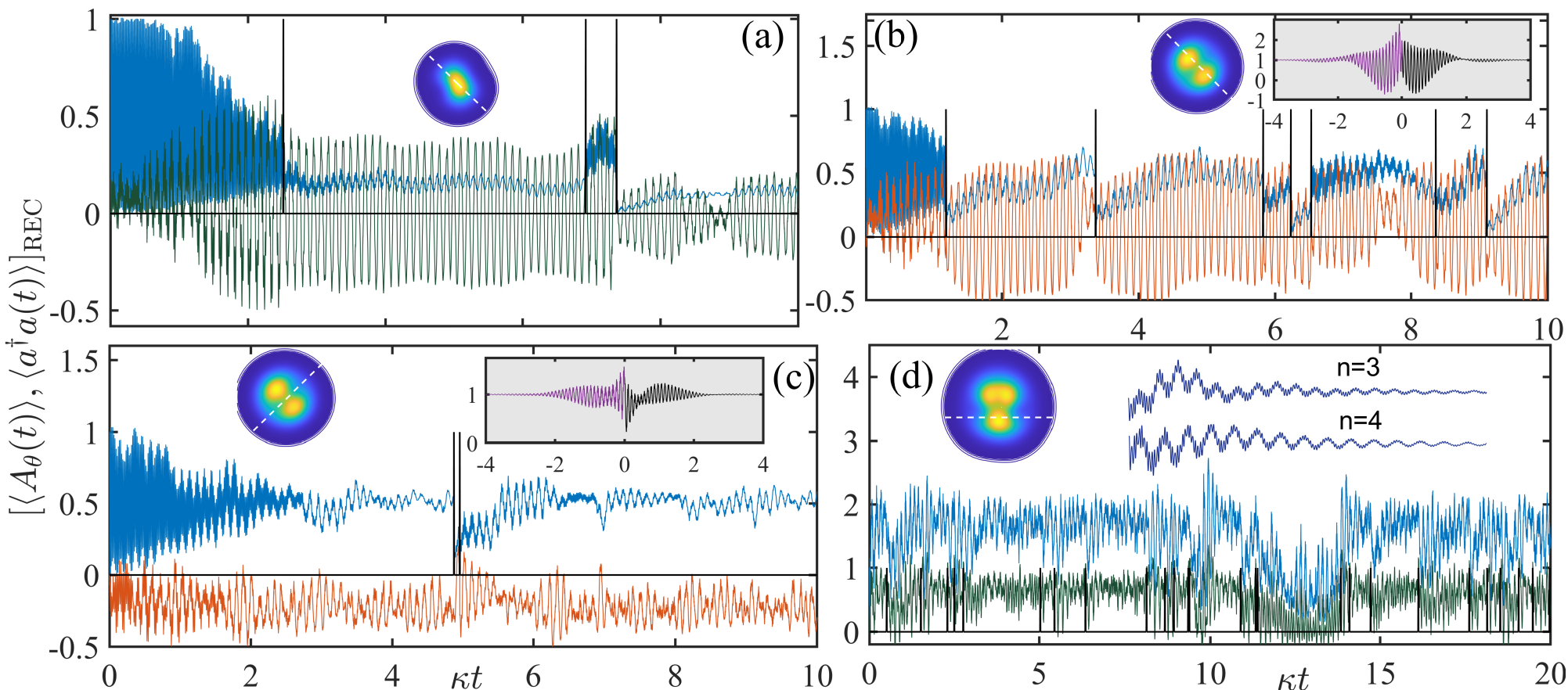}
 \caption{{\it Sampling a homodyne current under low-amplitude nonlinearity.} Individual trajectories of the conditioned intracavity photon number (positive-valued plots) and the conditioned field (plots with alternating sign), measured for $r=0.5$ with initial condition $\rho(0)=|1,-\rangle \langle 1, -|$ at different values of the driving frequency and strength $(\Delta\omega_d/g, \varepsilon_d/g)$: {\bf (a)} $(-0.7114, 0.03)$, {\bf (b, c)} $(-0.7114, 0.055)$ and {\bf (d)} $(0.545, 0.16)$. The inset contour plots in each frame depict the steady-state Wigner function, while the white dashed line indicates the direction set by the phase $\theta$ of the LO. The two additional insets in frame (d) depict the variation of the correlation function of the forward scattered photons $g_{\rm ss}^{(2)}(\tau)$ over an average photon lifetime, at the peaks of the three ($n=3$) and four ($n=4$) photon resonances for $\varepsilon_d/g=0.16$. The shaded insets of (b, c) depict unconditional wave-particle correlation functions $h_{\pi/4}(\tau)$ and $h_{3\pi/4}(\tau)$, respectively, normalized to unity at long delays. In all frames, $g/\kappa=200$ and $\gamma/(2\kappa)=0$. A MC algorithm unravelling the ME~\eqref{eq:ME1} under the action of the wave-particle correlator was developed {\it ad hoc} in \textsc{Matlab} using a basis truncated at the 14-photon level~\cite{KloedenPlaten}.}
 \label{fig:sampling}
\end{figure*}
 
From the concrete visualization offered by the quantum trajectory approach we gain significant understanding of the departure from the steady-state profile, underlying the dynamical unfolding of multi-photon resonances in an explicitly open quantum system setup. The complete picture is the complement of all the separate pictures -- an infinity of unravelings are obtained for different values of $r$ and $\theta$ -- and by the very nature of quantum mechanics no single picture can substitute for them all. In other words, the ME~\eqref{eq:ME1} carries the many pictures forward in parallel and the individual trajectories revealing the conditioned separation of timescales separate these pictures out~\cite{Carmichael1993}. 

For an initial guidance on $\theta$, we rely on the form of the steady-state Wigner function from the solution to ME~\eqref{eq:ME1}, and select the phase of the LO along those field quadratures which are characteristic to the development of nonlinearity as a function of the drive parameters $(\Delta\omega_d/g, \varepsilon_d/g)$. We first consider an unraveling along $\theta=3\pi/4$, probing a direction close to the anti-squeezed quadrature, for a weak drive prior to the onset of bimodality. Figure~\ref{fig:sampling}(a) shows a transient uninterrupted decay of the quantum beat before a first cavity emission marks the relaxation towards steady state, to be followed by a closely-spaced photon pair indicating photon bunching and reviving the quantum beat [we calculate $g_{\rm ss}^{(2)}(0)=2.17$]. When the first photon of a pair is emitted from the cavity, the conditioned mean photon number -- whence the collapse probability $2\kappa\langle\psi_c(t)| a^{\dagger}a|\psi_c(t)\rangle dt$ -- jumps upwards; this ensures that the second photon from the cascade will be emitted within a short time, here $\tau<(2\kappa)^{-1}$ after the first. The quantum beat on top of a segment of semiclassical evolution reappears as a feature of coherent superposition between these two jumps.

Increasing the drive amplitude we enter the regime of steady-state bimodality where $g^{(2)}_{\rm ss}(0)<1$. All but one of the seven photon emissions visible in Fig.~\ref{fig:sampling}(b) are associated with jumps downwards in $\langle a^{\dagger}a(t) \rangle_{\rm REC}$. After the initial interruption of the beat, the semiclassical oscillation converges to the two steady-state attractors, while quantum jumps bring it close to the unstable state near the phase-space origin. Contrary to what we expect from steady-state antibunching, aligning the LO in the orthogonal direction ($\theta=\pi/4$) to the development of bimodality, generated a sample trajectory with a single closely-spaced pair, stopping the relaxation to the steady state with $\langle a^{\dagger}a \rangle_{\rm ss}\approx 0.54$. At the same time, the unconditional cross-correlation of the intensity and the field amplitude of the forward-scattered light, $h_{\theta}(\tau)=\langle A_{\theta}(\tau)\rangle_c\equiv \langle a(\tau) \rangle_c e^{-i\theta} + {\rm c.c.}\, (\tau\geq 0)$, where $\langle a \rangle_c \equiv {\rm tr} [a\, e^{\mathcal{L}\tau}(a\rho_{\rm ss}a^{\dagger})]$, and its time-reversed version for $\tau\leq 0$~\cite{CarmichaelFosterChapter}, reveal an asymmetry of fluctuations~\cite{Denisov2002} depending on the chosen value for $\theta$. An asymmetric $h_{\theta}(\tau)$ rules out the validity of Gaussian statistics and signals the breakdown of detailed balance~\cite{Denisov2002} in the region of bimodality. A pronounced temporal asymmetry of intensity-field correlations before and after a trigger ``click'' has been identified for three-level atoms in Ref.~\cite{Xu2015}. For photon numbers $\sim 10^{-4}$ where we note the presence of strong bunching ($r=1$)~\cite{Shamailov2010} and a negative spectrum of squeezing ($r=0$), the correlation function attains very large negative values, violating its classical bounds as a result of the anomalous phase of the amplitude oscillation first noted for squeezed light in~\cite{GiantViolations2000}.

It is also interesting to note that the characteristic long span of the quantum beat (owing to the purposefully selected initial coherent-state superposition of the couplet states) for the two-photon resonance of Figs.~\ref{fig:sampling}(b, c) is replaced by a weaker transient interference of beats for the trimodality of Fig.~\ref{fig:sampling}(d), while the persistent semiclassical oscillations are replaced by a pattern reminiscent of traditional amplitude bistability involving two long-lived metastable states (to produce $\langle a^{\dagger}a \rangle_{\rm ss}\approx 0.88$ with a critical slowing down~\cite{Brookes2021, PhotonBlockade2015} in the average response) even when aligning the LO orthogonal to steady-state phase-space trimodality. The two inset plots of Fig.~\ref{fig:sampling}(d) testify to a reduced intensity correlation as higher-order resonances are accessed for the same value of $\varepsilon_d/g$; moreover, the photon stream is no longer antibunched at the four-photon resonance peak. 
\begin{figure*}
 \includegraphics[width=\textwidth]{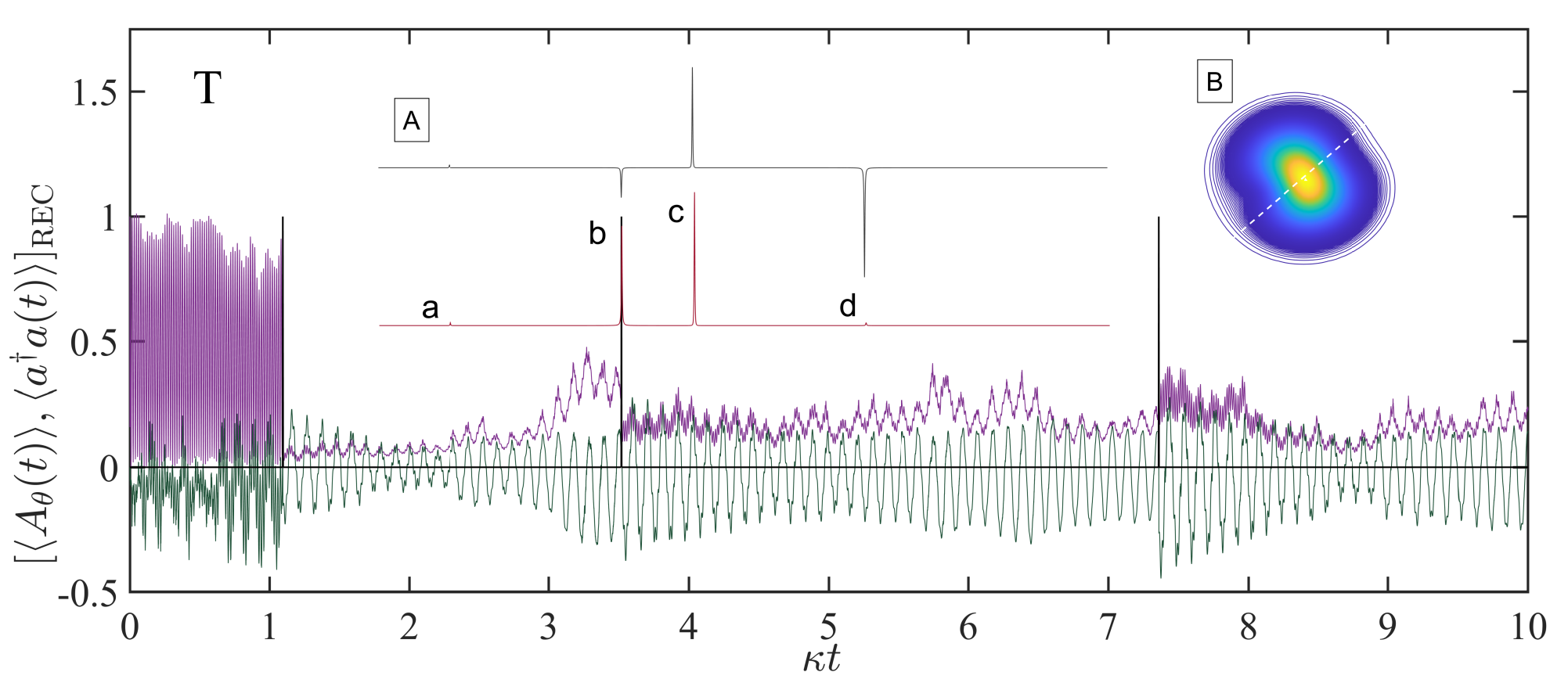}
  \includegraphics[width=0.45\textwidth]{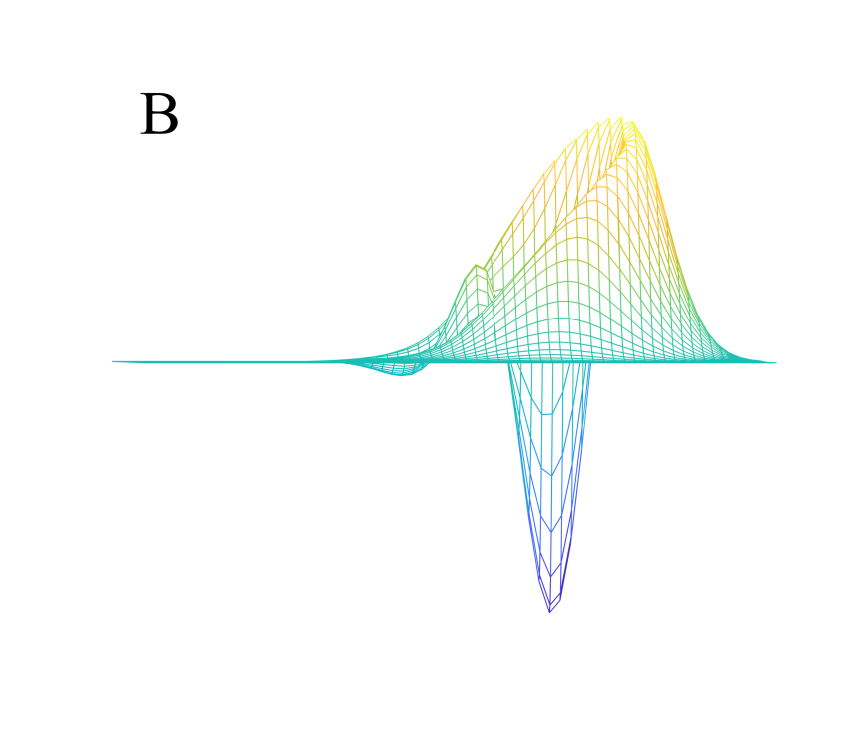}
   \includegraphics[width=0.43\textwidth]{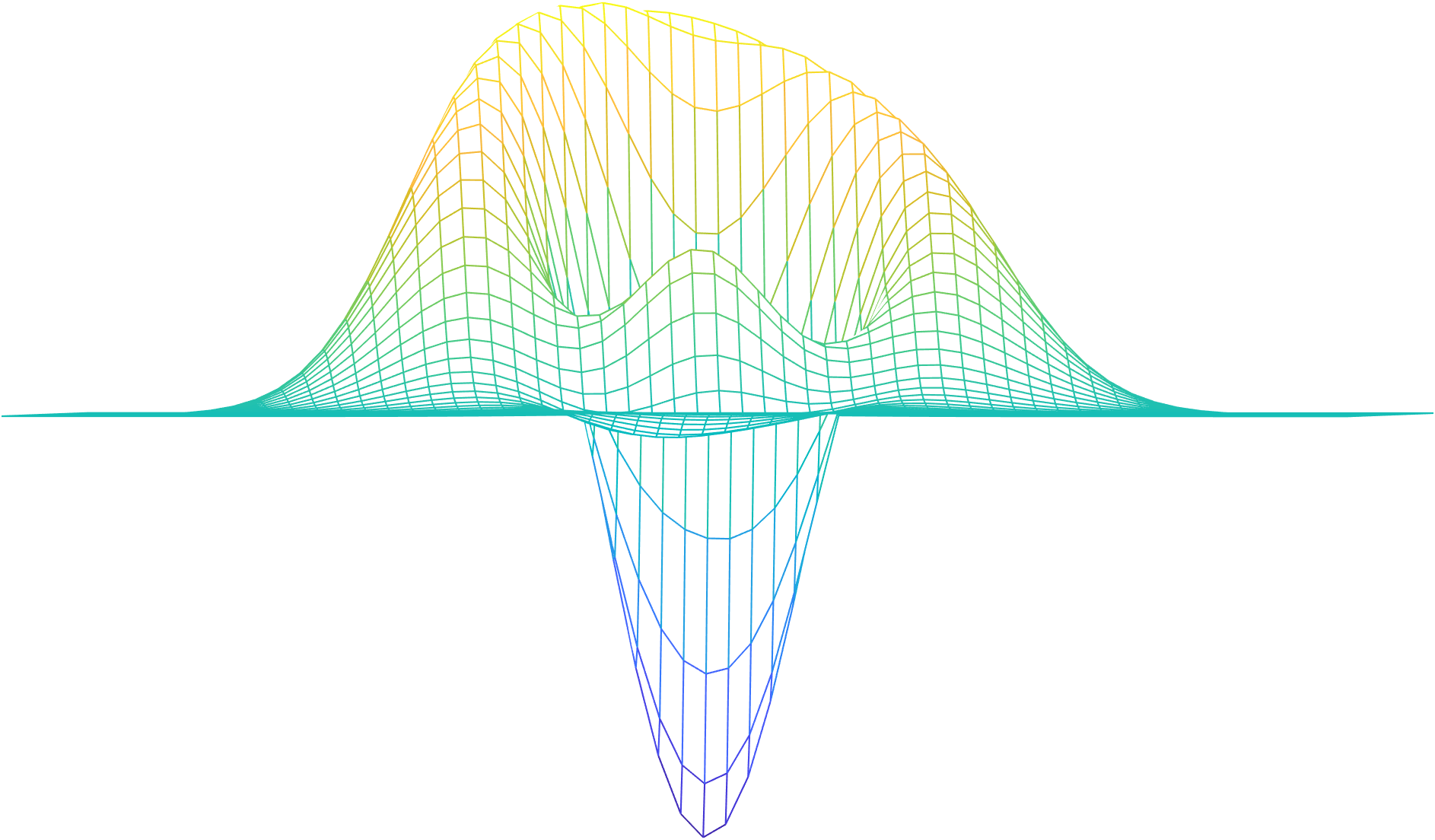}
 \caption{{\it Wave-particle correlator unraveling the ME in a direction orthogonal to steady-state bi(tri)modality.} ({\bf T}op) Individual trajectory of the conditioned intracavity photon number (positive-valued plot in blue) and the conditioned field (plot with alternating sign in orange) forming part of the homodyne photocurrent [see Eq.~\eqref{eq:di}] for $r=1/2$. The operation point is the same as in Fig.~\ref{fig:sampling}(a), except for the phase of the LO which is set to $\theta=\pi/4$. Inset A depicts the unlabeled spectrum of squeezing $\mathcal{S}_{\rightarrow}^{\theta}(\omega)$ against $(\omega-\omega_0)/\kappa$ (upper plot) on top of the transmission spectrum $T(\omega)$ (lower plot) calculated from Eq.~\eqref{eq:TSp} [the four peaks labeled $a, b, c, d$ correspond to transitions $|\xi_3 \rangle \to |\xi_2 \rangle$, $|\xi_1 \rangle \to |\xi_0 \rangle$, $|\xi_3 \rangle \to |\xi_1 \rangle$ and $|\xi_1 \rangle \to |\xi_0 \rangle$, respectively], while inset B depicts a contour plot of the steady-state Wigner function. ({\bf B}ottom) Schematic representation of the conditioned Wigner function (two projections) of the intracavity field at the time $\kappa t_{\rm max} \approx 9.95$ when the conditional photon number plotted in Fig.~\ref{fig:sampling}(d) reaches its maximum, $\langle a^{\dagger} a \rangle _{\rm REC, max} \approx 2.74$. The zero level is colored in green, midway between the main peak (yellow) and dip (blue).}
 \label{fig:orthogonal}
\end{figure*}

Furthermore, a few background comments are in order concerning the fluctuating conditioned field amplitude $\langle\psi_c(t)| A_{\theta}|\psi_c(t)\rangle$ plotted in all frames of Fig.~\ref{fig:sampling}, in conjunction with the photon number. This quantity is of particular importance since it forms part of the measured differential photocurrent together with the Gaussian white noise. There is a clear time asymmetry with respect to the origin set by each trigger photodetection at the APD. The BHD back action tends to recover the symmetry, albeit in a background of intense fluctuations. The back action comes about because of the quantum superposition of the LO and the JC source field. These two fields should not be thought of as being separable since each time a LO photon is detected the evolution of the cavity field is affected. This entails that the conditioned state $|\psi_c(t)\rangle$ at time $t$ is correlated with the shot noise at the detectors 1 and 2 from the recent past. Consequently, through the APD trigger clicks -- the particle aspect in the duality -- we postselect a subensemble of the shot noise which has been filtered through the dynamical response function of the source in the course of its coherent evolution.

As a further instance of the wave-particle unraveling (with $r=1/2$), the top frame of Fig.~\ref{fig:orthogonal} depicts an individual realization corresponding to a negative steady-state spectrum of squeezing (one obtained for $r=0$). Despite the presence of photon bunching [with $g_{\rm ss}^{(2)}(0)=2.17$ -- see Fig.~\ref{fig:g2s}(b)] the three photon emissions are spaced further apart than the inverse of the average photon lifetime $(2\kappa)^{-1}$. Interestingly, the same number of emissions occurs compared to the unraveling with $\theta=3\pi/4$ where the quadrature amplitude is larger [see Fig.~\ref{fig:sampling}(a)]. The steady-state cavity photon number calculated from the solution of the ME~\eqref{eq:ME1} is $\langle a^{\dagger}a \rangle_{\rm ss}\approx 0.16$. In conjunction with the photon emission probabilities, the conditioned Wigner function depicted in the bottom frame of Fig.~\ref{fig:orthogonal} -- corresponding to the maximum conditional emission probability plotted in Fig.~\ref{fig:sampling}(a) -- brings two aspects of the dynamics together: Quantum interference between dressed states produces a succession of peaks and dips with alternating signs as remnants of steady-state bistability, aligned in the direction imposed by the LO (set at $\theta=0$) -- once again, variations on the theme of conditionally generating single(few)-photon-added coherent states we met in Fig.~\ref{fig:particleWss}(a). In the steady-state distribution computed from the time-averaged density operator (the time average is equivalent to the ensemble average since quantum trajectories are ergodic~\cite{CarmichaelQO2}) the interference is interrupted at random times, dephasing to the purely positive profiles of Fig.~\ref{fig:sampling}.   

The BHD detector of bandwidth $\Gamma$ measures the fluctuations in the field amplitude -- the wave aspect in the duality -- before the arrival of another triggering photon~\cite{Reiner2001}. To recover the signal out of the residual shot noise $\xi(\tau)$ with correlation function $\overline{\xi(0)\xi(\tau)} \propto (\Gamma/N_s)\exp(-\Gamma \tau)$, the ensemble of $N_s \gg 1$ current samples initiated at the APD triggers $\{t_j\}$ is to be averaged as~\cite{GiantViolations2000, CarmichaelFosterChapter}$ \mathcal{H}(\tau)=(1/N_s)\sum_{j=1}^{N_s}i(t_j + \tau)$. Neglecting third-order moments and considering the limit of large detection bandwidth and negligible residual shot noise ($N_s \to \infty$), the average photocurrent $\mathcal{H}(\tau)$ is connected to the wave-particle correlation function $h_{\theta}(\tau)$ of the field fluctuations $\Delta a$ via the expression~\cite{Reiner2001} (for a non-zero steady-state field amplitude $\langle a \rangle_{\rm ss}$)
\begin{equation}\label{eq:Haverage}
\begin{aligned}
 \overline{h_{\theta}(\tau)}=&\lim_{N_s \to \infty} h_{\theta}(\tau)=\frac{\mathcal{H}(\tau)}{|\langle a \rangle_{\rm ss}| \sqrt{8\kappa (1-r)}}\\
 &=1+2\frac{\langle : \Delta A_{\theta}(0)\Delta A_{\theta}(\tau): \rangle_{\rm ss}}{|\langle a \rangle_{\rm ss}|^2 + \langle \Delta a^{\dagger} \Delta a \rangle_{\rm ss}},
 \end{aligned}
\end{equation}
where $\Delta A_\theta\equiv \frac{1}{2}(\Delta a e^{-i\theta}+\Delta a^{\dagger}e^{i\theta})$ is the quadrature amplitude fluctuation operator probed by the setting of the LO phase. From Eq.~\eqref{eq:Haverage} we find that assuming Gaussian statistics for a weak source field, the wave-particle correlation of the fluctuations $h_{\theta}(\tau)-1$ [see Eq.~\eqref{eq:Haverage}] and the spectrum of squeezing form a Fourier-transform pair, which means that small violations of nonclassicality in one part lead to large violations in the other (see~\cite{GiantViolations2000} and Sec. 2 of~\cite{CarmichaelFosterChapter}). On top of calculating unconditional quantum averages, we examine the regression of the fluctuations to steady state in the course of single trajectories generated in the wave-particle correlator unraveling. Note that the quantity $\langle : \Delta A_{\theta}(0)\Delta A_{\theta}(\tau): \rangle_{\rm ss}$ appearing in Eq.~\eqref{eq:Haverage} is equal to $R_{\theta}(\tau)$ introduced in Sec.~\ref{sec:waveasp}, since $\langle a\rangle_{\rm ss}=0$. As low strengths of the drive {\it operationally} qualify as those for which the detected forward photon emission stream is highly bunched~\cite{Shamailov2010}.

For the two-photon resonance and under the secular approximation underlying the minimal model, $\langle a \rangle_c=0$, since the matrix $(a\rho_{\rm ss}a^{\dagger})$ resulting after the emission of the ``first photon'' post steady state only contains diagonal matrix elements and the quantum beat, none of which are contained in the dressed-state resolution of $a$~\cite{Shamailov2010}. The fact that $\langle a \rangle_c=0$ for every quadrature phase amplitude of the field is also suggested by the circular symmetry characterizing the Wigner function of the state $\langle + |(a\rho_{\rm ss}a^{\dagger})|+ \rangle + \langle - |(a\rho_{\rm ss}a^{\dagger})|- \rangle$. As we have seen, in the general case $h_{\theta}(\tau)$ involves $\langle a \rangle_c \equiv {\rm tr} [a\, e^{\tilde{\mathcal{L}}\tau}(a\rho_{\rm ss}a^{\dagger})]$ and vanishes for $\tau \geq 0$, while for $\tau<0$ the unconditional wave-particle correlation involves the average ${\rm tr} [a^{\dagger}a e^{\tilde{\mathcal{L}}|\tau|}(a\rho_{\rm ss})]$ which evaluates to zero as well (without applying any coherent offset to the cavity output~\cite{GiantViolations2000}). The nonzero $h_{\theta}(\tau)$, in which the intermediate timescale dominates, and the asymmetric relaxation of the oscillations in the individual realizations of Figs.~\ref{fig:sampling}(b-c) testify to the departure of the quantum dynamics from the predictions of the minimal model. In Fig.~\ref{fig:sampling}(d), we also find a large positive cavity-field amplitude following the photon number in a pattern conforming to the conventional amplitude bistability. Moreover, the displayed time asymmetry in the fluctuations in principle precludes the connection of $h_{\theta}(\tau)$ with the spectrum of squeezing for very low strengths of the drive field exciting a JC multiphoton resonance, since autocorrelations are time-symmetric for a stationary process~\cite{Denisov2002, CarmichaelFosterChapter}. For JC resonances of higher order than second examined in this work, even when the source field is weak ($\varepsilon_d/\kappa \ll 1$), preliminary numerical results show that the unconditional wave-particle correlation function is once more not symmetric with respect to the time origin, the two inset plots in Figs.~\ref{fig:sampling}(b,c), with the asymmetry also being a function of $\theta$. 

\section{Conclusion}

We have demonstrated notable departures from the {\it unconditional} ME dynamics in the {\it conditional} photon counting time series generation as well as in the selection of the light amplitude by a nonclassical photon stream for the wave-particle correlator unraveling of a JC ME. The quantum beat -- a fast coherent oscillation unveiling the JC spectrum with its $\sqrt{n}$-nonlinearity -- is the common denominator in all realizations under multiphoton resonance operation. A conditional separation of timescales unfolds in the course of a particular unraveling dictated by the selected quadrature amplitude of the signal field. We have visualized the time-asymmetric fluctuations of the field amplitude triggered by recordable photon counting sequences, and have also explored the possibility of an operational determination~\cite{Lutterbach1997, Nogues2000, Lvovsky2001, Lvovsky2002, Bertet2002} of phase-space {\it quasi}probability distributions and photon-state superpositions~\cite{Resch2002A, Resch2002B,Abah2020} for the different unravelings. 

The dual nature of the measurement process, allowing the use of quantum trajectory theory to unravel the ME in two distinct ways, reappraises the BKS proposal of light-matter interaction by giving emphasis on the traceable coherent dynamics~\cite{CarmichaelTalk2000}. In our discussion, we have seen that it does so in the strong-coupling limit of light-matter interaction, one where photon blockade persists~\cite{SC2019}, meaning that the discrepancy between the mean-field predictions and the quantum dynamics continues to grow as the relevant system-size parameter $n_{\rm sc, s}=[g/(2\kappa)]^2$ is sent to infinity~\cite{PhotonBlockade2015}. In such a limit, fluctuations dominate over the steady-state amplitude, and the addition or subtraction of a few quanta -- those organizing the multiphoton resonances -- plays a significant role in the experimentally observable and contextual system response.

\begin{acknowledgments}
 I acknowledge financial support from the Severo Ochoa Center of Excellence as well as from the Swedish Research Council (VR).
\end{acknowledgments}

\appendix
\onecolumngrid

\section{Derivation of the waiting-time distribution}

To assess the corpuscular nature of the forward-scattered light, we employ the waiting-time distribution defined by conditional exclusive probability densities. This quantity is given by~\cite{Carmichael1989, Carmichael1993}
\begin{equation}
\begin{aligned}
 w_{{\rm ss},\,\rightarrow}(\tau)&=2\kappa \frac{{\rm tr}_S[a^{\dagger}a e^{\overline{\mathcal{L}} \tau}(a \rho_{\rm ss}a^{\dagger})]}{\braket{a^{\dagger}a}_{\rm ss}}\\
 &=(2\kappa)\,{\rm tr}_S\{a^{\dagger}a e^{\overline{\mathcal{L}} \tau} [\rho_{\rm cond}]\},
 \end{aligned}
\end{equation}
where $\overline{\mathcal{L}} \equiv \mathcal{L}-2\kappa a \cdot a^{\dagger}$ and $\rho_{\rm cond} \equiv (a \rho_{\rm ss} a^{\dagger})/\braket{a^{\dagger}a}_{\rm ss}$ is the (normalized) conditioned density matrix following the post-steady-state emission of the ``first'' photon. We remark that the intensity correlation function is instead given by~\cite{Carmichael1989, Carmichael1993} $g_{\rm ss}^{(2)}(\tau)={\rm tr}_S\{a^{\dagger}a e^{\mathcal{L} \tau} [\rho_{\rm cond}]\}/\braket{a^{\dagger}a}_{\rm ss}$ to reflect an unconditional probability (with a Liouvillian $\mathcal{L}$ instead of $\overline{\mathcal{L}}$).

We move now to the minimal model in the dressed-state basis, where $\mathcal{L}$ is replaced by $\tilde{\mathcal{L}}$. Then the vector $\boldsymbol{x}\equiv (\overline{\rho}_{00}, \overline{\rho}_{33}, {\rm Im}(\overline{\rho}_{03}))^{T}$ solves the homogeneous system of equations $\dot{\boldsymbol{x}}=\boldsymbol{M} \boldsymbol{x}$, where 
\begin{equation}
 \boldsymbol{M}=\begin{pmatrix}
0 & 0 & -2\Omega\\
0 & -3\kappa & 2\Omega \\
\Omega & -\Omega & -3\kappa/2
\end{pmatrix},
\end{equation}
with eigenvalues $\lambda_1=-3\kappa/2$ and $\lambda_{2(+), 3(-)}=\lambda_1 \pm \delta$, with $\delta\equiv |\lambda_1|\sqrt{1-16\Omega^2/(9\kappa^2)}$. The solution of the above system of equations can be written as
\begin{equation}
 \boldsymbol{x}(\tau)=\boldsymbol{S} e^{\boldsymbol{\Lambda}\tau} \boldsymbol{S}^{-1} \boldsymbol{x}(0),
\end{equation}
where $\boldsymbol{\Lambda}={\rm diag} (\lambda_{1}, \lambda_{2}, \lambda_{3})$, the columns of the matrix $\boldsymbol{S}$ are the right eigenvectors of $\boldsymbol{M}$ and the vector $\boldsymbol{x}(0)$ is populated by the elements of the conditioned density matrix. The only non-zero element is $[\boldsymbol{x}(0)]_{1}=[\rho_{\rm cond}]_{00}=\frac{1}{2}(p_1 + p_2)/(3p_3)=\frac{1}{2}$. Now, for the calculation of photon number under the action of $\overline{\mathcal{L}}$ only the matrix element $\boldsymbol{x}_{2}=\overline{\rho}_{33}(\tau)$ is of interest, for which we find
\begin{equation}
 \overline{\rho}_{33}(\tau)=\frac{2\Omega^2}{9\kappa^2-16\Omega^2}e^{-3\kappa \tau/2}\sinh^2\left(\frac{\tau \sqrt{9\kappa^2-9\Omega^2}}{4}\right).
\end{equation}
The rest of the matrix elements under the evolution with $\overline{\mathcal{L}}$ are uncoupled, and we find the simple expressions
\begin{equation}
 \overline{\rho}_{11}(\tau) + \overline{\rho}_{22}(\tau)=\tfrac{1}{3}\left[\left(\tfrac{\sqrt{2}+1}{2}\right)^2 + \left(\tfrac{\sqrt{2}-1}{2}\right)^2 \right]e^{-\kappa \tau}=\tfrac{1}{2}e^{-\kappa\tau}
\end{equation}
and
\begin{equation}
  \overline{\rho}_{12}(\tau)= \overline{\rho}_{21}^{*}(\tau)=\tfrac{1}{12}e^{-\kappa \tau}e^{i\nu \tau}.
\end{equation}
Finally, we express the photon-number operator in terms of the dressed states in the truncated Hilbert space as~\cite{Shamailov2010, Lledo2021}
\begin{equation}
 a^{\dagger}a\approx \tfrac{1}{2}\left(|1\rangle \langle 1|+|2\rangle \langle 2|+|2\rangle \langle 1| + |1\rangle \langle 2| \right) + \tfrac{3}{2} |3\rangle \langle 3|.
\end{equation}
Putting the pieces together we find
\begin{equation}
\begin{aligned}
  w_{{\rm ss}, \rightarrow}(\tau)&=2\kappa \left[\tfrac{1}{2}[\overline{\rho}_{11}(\tau) + \overline{\rho}_{22}(\tau) + 2{\rm Re}\{\overline{\rho}_{12}(\tau)\}]+\tfrac{3}{2}\overline{\rho}_{33}(\tau) \right]\\
  &=2\kappa\Bigg[\tfrac{1}{2}\left(\tfrac{1}{2}e^{-\kappa \tau} + \tfrac{1}{6}e^{-\kappa \tau}\cos(\nu \tau)\right)+\frac{3\Omega^2}{9\kappa^2-16\Omega^2}e^{-3\kappa \tau/2}\sinh^2\left(\frac{\tau \sqrt{9\kappa^2-9\Omega^2}}{4}\right)\Bigg].
  \end{aligned}
\end{equation}
The terms inside the small parentheses of the second lime, evaluated at the minima of the quantum beat define the lower asymptote $s_L(\tau)=\kappa(\frac{1}{2}e^{-\kappa\tau}-\frac{1}{6}e^{-\kappa \tau})=\kappa\frac{1}{3}e^{-\kappa\tau}$ plotted in Fig. 2(b). The upper asymptote $s_U(\tau)=\kappa(\frac{7}{12}e^{-\kappa\tau} + \frac{3}{2}e^{-3\kappa\tau/2})$ is obtained for $\Omega/\kappa \gg 1$.

\section{Derivation of the squeezing spectrum}

For the spectrum of squeezing we will employ the quantum regression formula to determine the following normal and time-ordered averages:
\begin{subequations}
\begin{align}
 \braket{a^{\dagger}(0)a(\tau)}_{\rm ss}&={\rm tr}_S \{a e^{\mathcal{L}\tau}[\rho_{\rm ss}a^{\dagger}]\},\label{eq:corr1}\\
  \braket{a(\tau)a(0)}_{\rm ss}&={\rm tr}_S \{a e^{\mathcal{L}\tau}[a\rho_{\rm ss}]\},\label{eq:corr2}
\end{align}
\end{subequations}
where ${\rm tr}_S$ indicates the trace taken over the system degrees of freedom. The equations of motion for the matrix elements involved in these two averages follow from the effective ME~\eqref{eq:ME2}. They can be divided into two autonomous pairs, with
\begin{subequations}\label{eq:sys1a}
\begin{align}
\dot{\rho}_{01}&=-\frac{\Gamma}{2} \rho_{01} -i \Omega\rho_{31},\\
\dot{\rho}_{31}&=-\frac{\Gamma_{31} + \Gamma_{32}}{2} \rho_{31} -i \Omega\rho_{01}
\end{align}
\end{subequations}
and the complex-conjugate elements,
\begin{subequations}\label{eq:sys1b}
\begin{align}
\dot{\rho}_{10}&=-\frac{\Gamma}{2} \rho_{10} +i \Omega\rho_{13},\\
\dot{\rho}_{13}&=-\frac{\Gamma_{31} + \Gamma_{32}}{2} \rho_{13} +i \Omega\rho_{10}
\end{align}
\end{subequations}
as well as 
\begin{subequations}\label{eq:sys2a}
\begin{align}
\dot{\rho}_{02}&=-\frac{\Gamma}{2} \rho_{02} -i \Omega\rho_{32},\\
\dot{\rho}_{32}&=-\frac{\Gamma_{31} + \Gamma_{32}}{2} \rho_{32} -i \Omega\rho_{02}
\end{align}
\end{subequations}
and the complex-conjugate elements,
\begin{subequations}\label{eq:sys2b}
\begin{align}
\dot{\rho}_{20}&=-\frac{\Gamma}{2} \rho_{20} +i \Omega\rho_{23},\\
\dot{\rho}_{23}&=-\frac{\Gamma_{31} + \Gamma_{32}}{2} \rho_{23} +i \Omega\rho_{20},
\end{align}
\end{subequations}
with initial conditions (at $\tau=0$) set by the quantum regression formula~\eqref{eq:corr1} as follows:
\begin{equation}\label{eq:incond}
\begin{aligned}
&{\rho}_{10}(0)=\frac{1}{\sqrt{2}}p_1, \quad {\rho}_{20}(0)=\frac{1}{\sqrt{2}}p_2,\\
&{\rho}_{01}(0)=\frac{\sqrt{2}+1}{2} \rho_{03, {\rm ss}}, \quad {\rho}_{02}(0)=\frac{\sqrt{2}-1}{2} \rho_{03, {\rm ss}},\\
&{\rho}_{31}(0)=\frac{\sqrt{2}+1}{2} p_3,  \quad {\rho}_{32}(0)=\frac{\sqrt{2}-1}{2} p_3,\\
&{\rho}_{13}(0)= \rho_{23}(0)=0,
\end{aligned}
\end{equation}
where $p_3=4\Omega^2/(9\kappa^2 + 20\Omega^2)$ is the steady-state excitation probability of $|\xi_3 \rangle$. Owing to the detailed balance established in the steady-state, the two intermediate levels $|\xi_1 \rangle$, $|\xi_2 \rangle$ are occupied with probabilities $p_1=(\Gamma_{31}/\Gamma)p_3$ and $p_2=(\Gamma_{32}/\Gamma)p_3$, respectively. The steady-state ``polarization'' in the subspace of the driven transition is calculated as $\rho_{03, {\rm ss}}=i 3\kappa p_3/(2\Omega)$.  

For a decay via the cavity channel, setting $\gamma=0$, we Laplace-transform the two sets of equations to find
\begin{equation}
{\rho}_{(1,2)0}(\tau)=\frac{p_{(1,2)}}{\sqrt{2}}e^{-\kappa |\tau|}\left[\cos(\mu \tau) + \frac{\kappa}{2\mu}\sin(\mu |\tau|)\right],
\end{equation}
\begin{equation}
{\rho}_{3(1,2)}(\tau)=\frac{\sqrt{2}\pm 1}{2}p_3 e^{-\kappa |\tau|}\left[\cos(\mu \tau) + d_{(1,2)}\sin(\mu |\tau|)\right],
\end{equation}
with 
\begin{equation}
d_{(1,2)}=-\frac{\frac{\kappa}{2}p_3+i\Omega p_{03,{\rm ss}}}{\mu p_3}=\frac{\kappa}{\mu},
\end{equation}
where $\mu \equiv \sqrt{\Omega^2-\kappa^2/4}$.

The required first-order correlation function then reads
\begin{equation}
\begin{aligned}
\braket{a^{\dagger}(0)a(\tau)}_{\rm ss}=&\frac{1}{\sqrt{2}}[{\rho}_{10}(\tau)+ {\rho}_{20}(\tau)]\\
& +\frac{\sqrt{2}+1}{2} {\rho}_{31}(\tau)+ \frac{\sqrt{2}-1}{2} {\rho}_{32}(\tau),
\end{aligned}
\end{equation}
while $\braket{a^{\dagger}(0)a(\tau)}_{\rm ss}=\braket{a^{\dagger}(\tau)a(0)}_{\rm ss}^{*}$. For the correlator $\braket{a(\tau)a(0)}_{\rm ss}$ we now find
\begin{equation}
\braket{a(\tau)a(0)}_{\rm ss}=\frac{1}{\sqrt{2}}[{\rho}_{10}(\tau)+ {\rho}_{20}(\tau)],
\end{equation}
where the matrix elements ${\rho}_{10}$ and ${\rho}_{20}$ satisfy the systems of Equations~\eqref{eq:sys1b} and~\eqref{eq:sys2b}, but now with initial conditions set by the quantum regression formula~\eqref{eq:corr2},
\begin{equation}
{\rho}_{(1,2)0}(0)=\frac{\sqrt{2}\pm 1}{2}\rho_{30, {\rm ss}}, \quad {\rho}_{(1,2)3}(0)=0.
\end{equation}
This yields the simpler expressions
\begin{equation}
{\rho}_{(1,2)0}(\tau)=\frac{\sqrt{2}\pm 1}{2}\rho_{30, {\rm ss}}e^{-\kappa |\tau|} \left[\cos(\mu \tau) + \frac{\kappa}{2\mu}\sin(\mu |\tau|) \right].
\end{equation}

We now consider decay via both channels with $\gamma=2\kappa$. The correlator $\braket{a^{\dagger}(0)a(\tau)}_{\rm ss}$ involves instead the matrix elements
\begin{equation}
{\rho}_{(1,2)0}(\tau)=\frac{p_{(1,2)}}{\sqrt{2}}e^{-(3\gamma/4)|\tau|}\left[\cos(\mu^{\prime} \tau) +\frac{\gamma}{4\mu^{\prime}} \sin(\mu^{\prime} |\tau|)\right],
\end{equation}
and
\begin{equation}
{\rho}_{3(1,2)}(\tau)=\frac{\sqrt{2}\pm 1}{2}p_3 e^{-(3\gamma/4) |\tau|}\left[\cos(\mu^{\prime} \tau) + d_{(1,2)}^{\prime}\sin(\mu^{\prime} |\tau|)\right],
\end{equation}
with 
\begin{equation}
d_{(1,2)}^{\prime}=-\frac{\frac{\gamma}{4}p_3+i\Omega p_{03,{\rm ss}}}{\mu^{\prime} p_3},
\end{equation}
where $\mu^{\prime}=\sqrt{\Omega^2-\gamma^2/16}$. As for the correlator $\braket{a(\tau)a(0)}_{\rm ss}$, we have
\begin{equation}
{\rho}_{(1,2)0}(\tau)=\frac{\sqrt{2}\pm 1}{2}\rho_{03, {\rm ss}} e^{-(3\gamma/4) |\tau|} \left[\cos(\mu^{\prime} \tau) + \frac{\gamma}{4\mu^{\prime}}\sin(\mu^{\prime} |\tau|)\right].
\end{equation}
For the spectrum of squeezing we need to calculate the normal and time-ordered variance:
\begin{equation}
R_{\theta}(\tau)\equiv \lim_{t\to \infty}\braket{:A_{\theta}(t) A_{\theta} (t+\tau):} \equiv \braket{:A_{\theta}(0) A_{\theta} (\tau):}_{\rm ss},
\end{equation}
where $A_{\theta}\equiv [a \exp(-i\theta) + a^{\dagger}\exp(i\theta)]/2$ and $\braket{:\,:}$ denotes normal ordering. This gives:
\begin{equation}
\begin{aligned}
4 R_{\theta}(\tau)&=\exp(-2i\theta)\braket{a(\tau)a(0)}_{\rm ss} + \braket{a^{\dagger}(\tau)a(0)}_{\rm ss}\\
&\quad + \braket{a^{\dagger}(0)a(\tau)}_{\rm ss} + \exp(2i\theta)\braket{a^{\dagger}(0)a^{\dagger}(\tau)}_{\rm ss}\\
&=2 {\rm Re}[\exp(-2i\theta)\braket{a(\tau)a(0)}_{\rm ss} + \braket{a^{\dagger}(0)a(\tau)}_{\rm ss}].
\end{aligned}
\end{equation} 
The normalized transmission spectrum is given by a sum of Laplace Transforms, 
\begin{equation}\label{eq:TSp}
T(\omega)=\frac{1}{\pi}\sum_{i,j}{\rm Re}\{\mathcal{P}_{ij}(\tilde{s}_{ij})\},
\end{equation}
with $\tilde{s}_{ij}=-i[(\omega-\omega_0+\tilde{E}_j/\hbar-\tilde{E}_{i}/\hbar)]/\kappa$, while the spectrum of squeezing in the forward direction reads
\begin{equation}\label{eq:spsq}
\begin{aligned}
\mathcal{S}_{\rightarrow}^{\theta}(\omega)&=2 (2\kappa) \int_0^{\infty}d\tau \cos(\omega \tau) R_{\theta}(\tau)\\
&= (4\kappa){\rm Re}\left\{\int_0^{\infty}d\tau e^{i\omega \tau} R_{\theta}(\tau)\right\}.
\end{aligned}
\end{equation}
We express the result in terms of linear combination of the general integral
\begin{equation}
I(\omega; a_1, b_1, \lambda)=\int_0^{\infty}dt\, e^{i\omega t} e^{-a_1 t} \left[\cos(b_1 t) + \lambda \frac{a_1}{2b} \sin(b_1 t)\right],
\end{equation}
distinguishing the cases where $b_1=\sqrt{\Omega^2-a_1^2/4}>0$ and $b_1= i|b_1|=i\sqrt{a_1^2/4-\Omega^2}$. In the former case, we find
\begin{equation}
\begin{aligned}
I=&\frac{1}{a_1^2+(\omega+b_1)^2}\left\{\frac{1}{2}\left[a_1 \left(1 + \lambda\frac{\omega+b_1}{2b}\right) + i\left(\omega + b_1 - \lambda\frac{a_1^2}{2b}\right)\right] \right\}\\
&+\frac{1}{a_1^2+(\omega-b_1)^2}\left\{\frac{1}{2}\left[a_1 \left(1 + \lambda\frac{b_1-\omega}{2b}\right) + i\left(\omega - b_1 + \lambda\frac{a_1^2}{2b}\right)\right] \right\}
\end{aligned}
\end{equation}
and in the latter,
\begin{equation}
\begin{aligned}
I=&\frac{1}{\omega^2+(a_1+|b_1|)^2}\left\{\frac{1}{2}\left[a_1 \left(1 + \frac{|b_1|}{a_1}-\lambda\frac{1}{2|b_1|}(a_1+|b_1|)\right) + i\omega\left(1 - \lambda\frac{a_1}{2|b_1|}\right)\right] \right\}\\
&+\frac{1}{\omega^2+(a_1-|b_1|)^2}\left\{\frac{1}{2}\left[a_1 \left(1 - \frac{|b_1|}{a_1}+\lambda\frac{1}{2|b_1|}(a_1-|b_1|)\right) + i\omega\left(1 +\lambda \frac{\kappa}{2|b_1|}\right)\right] \right\}.
\end{aligned}
\end{equation}
\twocolumngrid

\bibliography{bibliography}

\end{document}